\documentstyle[aps,prl,multicol,epsf]{revtex}
\begin{document}
\title{\bf Steady State of an Inhibitory Neural Network}
\author{P.~L.~Krapivsky and S.~Redner}

\address{Center for BioDynamics, Center for Polymer Studies,
and Department of Physics, Boston University, Boston, MA, 02215}
\maketitle
\begin{abstract}

\noindent
We investigate the dynamics of a neural network where each neuron evolves
according to the combined effects of deterministic integrate-and-fire
dynamics and purely inhibitory coupling with $K$ randomly-chosen
``neighbors''.  The inhibition reduces the voltage of a given neuron by an
amount $\Delta$ when one of its neighbors fires.  The interplay between the
integration and inhibition leads to a steady state which is determined by
solving the rate equations for the neuronal voltage distribution.  We also
study the evolution of a single neuron and find that the mean lifetime
between firing events equals $1+K\Delta$ and that the probability that a
neuron has not yet fired decays exponentially with time.

\medskip\noindent {PACS numbers: 87.18.Sn, 87.19.La, 07.05.Mh.}
\end{abstract}

\begin{multicols}{2}
%%%%%%%%%%%%%%%%%%%%%%%%%%%%%%%%%%%%%%%%%%%%%%%%%%%%%%%%%%%%%%%%%%%%%%

\section{Introduction}

Networks of neurons which undergo ``spiky'' dynamics have been thoroughly
investigated (see e.g.\cite{brunel,gh} and references therein).
Nevertheless, a theory which describes the dynamics of randomly
interconnected excitatory and inhibitory spiking neurons is still lacking.
Even a system composed exclusively of inhibitory
neurons\cite{va,gr,wb,white,bh} appears too complicated for analytical
approaches.  Part of the reason for this is that the dynamics of a single
neuron involves many physiological features across a wide range of time
scales that are difficult to incorporate into an analytical theory.  Our goal
in this work is to describe some of the dynamical features of a purely
inhibitory neural network within the framework of a minimalist model.  While
we sacrifice realism by this approach, our model is analytically tractable.
This feature offers the possibility that more realistic networks may be
treated by natural extensions of our general framework.

We specifically investigate an integrate-and-fire neural network, in which
the integration step is purely linear in time, and in which there exists only
inhibitory and instantaneous coupling between interacting neurons.  We thus
ignore potentially important features such as voltage leakage during the
integration, as well as heterogeneity in the external drive and in the
network couplings.  However, we do {\em not\/} assume all-to-all coupling, an
unrealistic construction which is often invoked as a simplifying assumption.
Instead, the (average) number of ``neighbors'' is a basic parameter of our
model.  Analytically, we consider {\em annealed} coupling, where the
neighbors of a neuron are reassigned after each neuron firing event.
However, our simulations indicate that the model with {\em quenched}
coupling, where the neighbors of each neuron are fixed for all time, gives
nearly identical results.

\begin{figure}
  \narrowtext \epsfxsize=2.3in\hskip 0.2in\epsfbox{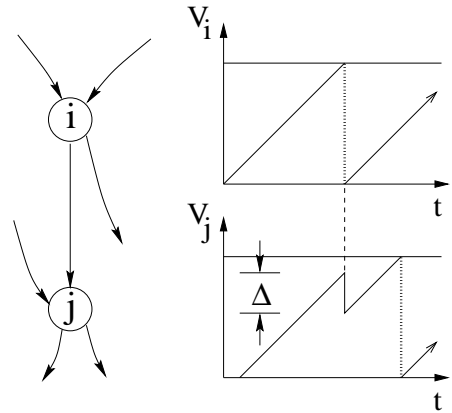} \vskip
  0.15in
\caption{Illustration of the model dynamics.  A single neuron $i$ undergoes
  deterministic integrate-and-fire dynamics (upper right).  When this neuron
  fires, its voltage $V_i$ is set to zero, while simultaneously the voltages
  on all its inhibitory-coupled neighbors are reduced by $\Delta$ (lower
  right).
\label{IF}}
\end{figure}

Each integrate-and-fire neuron is represented by a single variable -- the
polarization level, or voltage $V$.  Our model has two fundamental
ingredients: (i) the dynamics of individual neurons, and (ii) the interaction
between them.  For the former, we employ deterministic integrate-and-fire
dynamics, in which the voltage on a single neuron increases linearly with
time until a specified threshold is reached\cite{integrate}.  At this point
the neuron suddenly fires by emitting an action potential and the neuronal
voltage quickly returns to a reference level (Fig.~\ref{IF}).  For
concreteness and without loss of generality, we assume the rate of voltage
increase and the threshold voltage are both equal to 1.  We also assume that
$V$ is instantaneously set to zero after firing, {\it i.\ e.}, we neglect
delays in signal transmission between neurons\cite{delay}.

The meaning of the inhibition is illustrated in Fig.~\ref{IF}.  When a given
neuron fires, it instantaneously transmits an inhibitory action potential to
$K$ randomly-chosen neighbors whose voltages are each reduced by an amount
$\Delta$.  This inhibition delays the time until these inhibited neurons can
reach threshold and ultimately fire themselves.  The neighbors of a given
neuron are selected at random from among all the neurons in the network and
they are chosen anew every time any neuron fires.  Thus the coupling in the
network is annealed.  While a network with fixed quenched coupling is
biologically much more realistic, annealed coupling means that a rate
equation provides the exact description of the dynamics.  Fortunately, the
annealed and quenched systems appear to be statistically identical when the
number of inhibitory-coupled neurons is sufficiently large.  We shall
consider only this limit in what follows.

Within the rate equation approach, the distribution of neuronal voltages in
our inhibitory network is described by linear dynamics except at the isolated
times when a neuron fires.  The underlying rate equation admits a steady
state voltage distribution whose basic properties are established
analytically in Sec.\ II.  In general, although the network has a steady
response, the dynamics of an individual neuron has an interesting history
between firing events.  We study, in particular, the probability that a
neuron ``survives'' up to a time $t$ after its last firing event in Sec.\ 
III.  The survival probability decays exponentially in time with a decay rate
that depends on the competition between the integration and the inhibitory
coupling.  This provides a relatively complete picture of the dynamics of a
single neuron in the network.  A summary is given in Sec.\ IV and some
identities are proven in the Appendix.

\section{Rate Equations and the Steady State}

For the rate equation description we assume that the number of neurons is
large and thus a continuum approach is appropriate.  We define $P(V,t)dV$ as
the fraction of neurons whose voltage lies in the interval $(V,V+dV)$.  Then
the probability density $P(V,t)$ obeys the master equation,
\begin{eqnarray}
\label{master}
\left({\partial\over \partial t}+{\partial\over \partial V}\right)
P(V,t)&=&P_1(t)\,\delta(V)\\
&+&KP_1(t)\left[P(V+\Delta,t)-P(V,t)\right],\nonumber
\end{eqnarray}
where $P_1(t)\equiv P(V=1,t)$.  The second term on left-hand side accounts
for the voltage increase because of the deterministic integration.  The first
term on the right-hand side accounts for the increase of zero-voltage neurons
due to the firing of other neurons which have reached the threshold value
$V_{\rm max}=1$.  The second set of terms accounts for the change in $P(V)$
due to the processes where $V+\Delta\to V$ and $V\to V-\Delta$ (we assume
$\Delta>0$ since inhibitory neurons are being considered).

In the steady state, Eq.~(\ref{master}) simplifies to
\begin{equation}
\label{PV}
{dP(V)\over dV}=P_1\,\delta(V)+KP_1\left[P(V+\Delta)-P(V)\right].
\end{equation}
Equation (\ref{PV}) may be easily solved by introducing the Laplace transform,
\begin{equation}
\label{Lap}
{\cal P}(s)=\int_{-\infty}^1 dV\,P(V)\,e^{sV}, 
\end{equation}
to give, after some straightforward steps, 
\begin{equation}
\label{Ps}
{\cal P}(s)={e^s-1\over K\left(e^{-s\Delta}-1\right)+s/P_1}.
\end{equation}
The unconventional definition of the Laplace transform reflects that fact that
the voltage is restricted to lie in the range $[-\infty,1]$.  

To solve for the Laplace transform, we first note that ${\cal P}(0)=1$ due to
normalization.  Combining this with Eq.~(\ref{Ps}), we obtain
$P_1=(1+K\Delta)^{-1}$ thus completing the solution.  The final result is
\begin{equation}
\label{Psol}
{\cal P}(s)={e^s-1\over K\left(e^{-s\Delta}-1\right)+s(1+K\Delta)}.
\end{equation}
It may be verified by elementary means that this function has a simple pole
at $s=-\lambda$, that is, ${\cal P}(s)={A/(s+\lambda)}+\ldots$, where
$\lambda$ is the root of
\begin{equation}
\label{root}
K\left(e^{\lambda\Delta}-1\right)-\lambda(1+K\Delta)=0, 
\end{equation}
and $A=(1-e^{-\lambda})/[\Delta(1+K\Delta)\lambda-1]$.  

The existence of a simple pole in the Laplace transform implies that the
voltage distribution itself has an exponential asymptotic tail as 
$V\to -\infty$, 
\begin{equation}
\label{Vinf}
P(V)\to A e^{\lambda V}.
\end{equation}
The limiting behaviors of the decay constant $\lambda$ may also be found from
Eq.~(\ref{root}) and give
\begin{equation}
\label{lim}
\lambda\to\cases{
\Delta^{-1}\ln\left[(K\Delta^{-1})\right]  &when $\Delta\to 0$,\cr
2(K\Delta^2)^{-1}                       &when $\Delta\to \infty$.\cr}
\end{equation}

\begin{figure}
  \narrowtext \epsfxsize=2.55in\hskip 0.25in\epsfbox{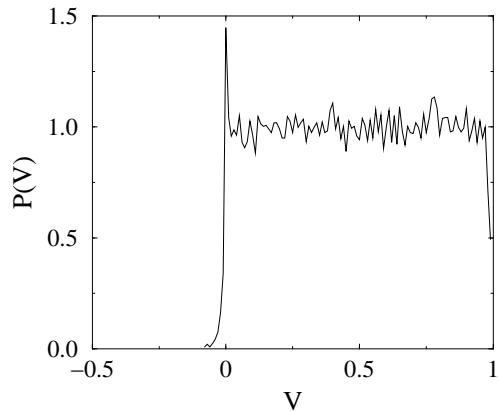} \vskip
  0.25in
\caption{Typical simulation results for the steady state neuronal voltage 
  distribution for a network of 25000 neurons, with annealed coupling to
  $K=50$ other neurons and $\Delta=1/50$.  The distribution is shown at 10
  time steps.
\label{sim}}
\end{figure}

To visualize these results, we have performed Monte Carlo simulations of an
inhibitory neural network whose dynamics is defined by Eq.~(\ref{master}).  A
representative result is shown in Fig.~\ref{sim} initial voltages uniformly
distributed in $[0,1]$.  Essentially identical results are obtained for
quenched inter-neuron coupling but all other system parameters unchanged.
After approximately one time unit, the system has reached the steady state
shown in the figure.  Other initial conditions merely have different time
delays until the steady state is reached.  In addition to the exponential
decay of the distribution for the smallest voltages, there are several other
noteworthy features.  First, there is a linear decay of the distribution to
its limiting value $P_1=(1+K\Delta)^{-1}$ as $V\to 1$ from below; this is
reflective of the absorbing boundary condition at $V=1$.  There is also a
sharp peak at $V=0$, corresponding to the continuous input of reset neurons.
For small $\Delta$, almost all voltages lie within the range $[0,1]$.  This
feature is reminiscent of the Bak-Sneppen evolution model\cite{bs} in which
the ``fitnesses'' of most species lie within a finite target range, together
with a small population of sub-threshold species.

Finally we note that Eq.~(\ref{master}) applies even when the number of
neighbors for a given neuron is not fixed.  In this case, $K$ may be
interpreted as just the {\em average} number of interacting neighbors of a
given neuron.  Therefore $K$ can be any positive real number.  The voltage
offset $\Delta$ can also be heterogeneous, {\it e.\ g.}, distributed in a
finite range with some density $\rho(\Delta)$.  With this generalization, the
term $P(V+\Delta)$ in Eq.~(\ref{PV}) should be replaced by $\int
d\Delta\,\rho(\Delta)P(V+\Delta)$.  The resulting equation is still solvable
by the same Laplace transform technique as in the homogeneous network and we
now obtain a similar solution to that in Eq.~(\ref{Psol}) but with
$P_1=(1+K\langle\Delta\rangle)^{-1}$, where $\langle\Delta\rangle=\int
d\Delta\,\rho(\Delta)\,\Delta$.

\section{Evolution of a Single Neuron}

\begin{figure}
  \narrowtext \epsfxsize=2.35in\hskip 0.4in\epsfbox{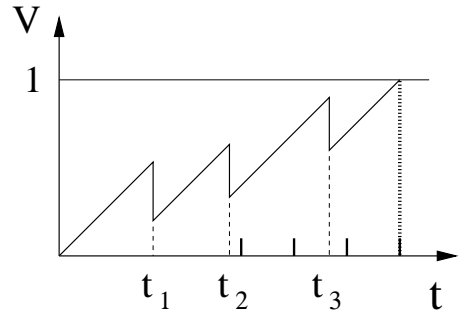} \vskip
  0.25in
\caption{Voltage trajectory of a typical neuron following a voltage reset.
  Because each inhibitory spike reduces the voltage by $\Delta$, a neuron
  which receives $k$ inputs fires exactly at time $\tau=1+k\Delta$.  The
  horizontal tick marks are at $t=1,1+\Delta,1+2\Delta$, and $1+3\Delta$.
\label{tagged}}
\end{figure}

In addition to the steady-state voltage distribution $P(V)$, we study the
time dependence of an individual ``tagged'' neuron in the steady state.
Consider, for example, a neuron which last fired at time $T_0$ which we set
to 0 for simplicity.  The fate of this neuron may be generally characterized
by its survival probability $S(t)$, defined as the probability that this
neuron has not yet fired again during the time interval $(T_0,T_0+t)$,
irrespective of how many inhibitory inputs it may have received
(Fig.~\ref{tagged}).  A more comprehensive description is provided by
$Q_k(t)$, the probability that the tagged neuron has not yet fired and that
it received $k$ inhibitory inputs during $(0,t)$.  Clearly, $S(t)=\sum_{k\geq
  0}Q_k(t)$.  As we now show, the survival probability and $Q_k(t)$ exhibit
non-trivial first-passage characteristics.

Before discussing the survival probability in detail, let us first understand
why a tagged neuron must eventually fire and why the survival probability
decays exponentially at long times.  In the absence of interactions, the
voltage of a neuron increases deterministically with rate 1.  On the other
hand, inhibitory spikes, each of which reduce the voltage by $\Delta$, occur
stochastically with rate $r=KP_1=K/(1+K\Delta)$.  This gives
$r\Delta={K\Delta/(1+K\Delta)}<1$ for the rate at which the voltage decreases
due to inhibition.  Thus, on average the voltage increases at a {\em net\/}
rate $1-r\Delta=(1+K\Delta)^{-1}$.  Consequently the voltage of a tagged
neuron must eventually reach the threshold and fire.

\begin{figure}
  \narrowtext \epsfxsize=3.0in\hskip 0.0in\epsfbox{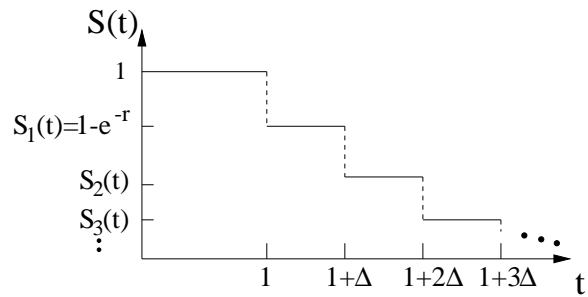} \vskip
  0.25in
\caption{Behavior of the survival probability as a function of time.  
  This function is constant within each interval $[1+(k-1)\Delta,1+k\Delta]$
  but exhibits an overall exponential decay in time.
\label{surv}}
\end{figure}

{}From the above argument, the mean lifetime $\langle t\rangle$ of a neuron
between firing events is just the inverse of this rate.  Thus
\begin{equation}
\label{avt}
\langle t\rangle=1+K\Delta. 
\end{equation}
Notice that the density of neurons which are firing, $P_1$, equals the
inverse lifetime, as expected naively.  Since the evolution of the neuronal
voltage is a random-walk-like process which is biased towards an absorbing
boundary in a semi-infinite geometry, the neuron survival probability must
decay exponentially with time\cite{redner},
\begin{equation}
\label{exp}
S(t)\propto e^{-t/\tau}\quad {\rm when} \quad t\to\infty.
\end{equation}

To understand the asymptotic behavior of $S(t)$, it is helpful to consider
first the voltage evolution of a single neuron which has experienced $0, 1,
2,\ldots$ inhibitory spikes.  It is important to appreciate that a neuron
which receives exactly $k$ spikes necessarily fires exactly at time
$1+k\Delta$ (Fig.~\ref{tagged}).  Thus the survival probability is a
piecewise constant function which changes discontinuously at $t_k=1+k\Delta$,
$k=0,1,2,\ldots$.

We now determine the value of $S(t)$ at each of these plateaux.  When $t<1$,
the tagged neuron has no possibility of firing and $S(t)\equiv S_0=1$.  To
survive until time $t=1+\Delta$, the neuron must receive at least one spike
in the time interval $[0,1]$.  Since neurons experience spikes at constant
rate $r=KP_1$, the probability of the neuron receiving no spikes in $[0,1]$
is $e^{-r}$.  The survival probability for $1<t<1+\Delta$ thus equals
$S(t)\equiv S_1=1-e^{-r}$ (Fig.~\ref{surv}).  Similarly, to survive until
time $1+2\Delta$, the neuron must receive one spike before $t=1$ and a second
spike before $t=1+\Delta$.  By writing the probabilities for each of these
events, we find, after straightforward steps,
\begin{equation}
\label{S2}
S_2= 1-e^{-r}-r\,e^{-r(1+\Delta)},\quad 1+\Delta<t<1+2\Delta.
\end{equation}
This direct approach becomes increasingly unwieldy for large times, however,
and we now present a more systematic approach.

To this end, we first solve for $Q_k(t)$, the probability that the tagged
neuron has experienced $k$ inhibitory inputs but has not yet fired.  For a
constant rate $r$ of inhibitory spikes, the probability that the tagged
neuron has been spiked exactly $k$ times, with each spike occurring in the
time intervals $[t_j,t_j+dt_j]$, $j=1,2,\ldots,k$, equals $e^{-rt}
\prod_{1\leq j\leq k} r\,dt_j$.  Therefore,
\begin{equation} 
\label{Qk}
Q_k(t)=r^k\,e^{-rt}\,\theta(1+k\Delta-t)\int \prod_{j=1}^k dt_j.
\end{equation}
Here the step function $\theta(1+k\Delta-t)$ guarantees that the voltage of
the tagged neuron is below the threshold at time $t$.  We must also ensure
that the voltage is less than one throughout the entire time interval
$(0,t)$.  The necessary and sufficient condition for this to occur is that
the voltage is below threshold at each spike event.  This determines the
integration range in Eq.~(\ref{Qk}) to be
\begin{equation}
\label{range}
t_{j-1}<t_j<{\rm min}[t,1+(j-1)\Delta]
\end{equation}
for $j=1,\ldots,k$ (we set $t_0\equiv 0$). 

We now evaluate $Q_k(t)$ successively for each time window
$[1+(j-1)\Delta,1+j\Delta]$.  Consider first the range $1<t<1+\Delta$.  Here
a tagged neuron which has not fired must have received at least one spike.
Consequently $Q_0(t)=0$.  Similarly, if a neuron receives a single spike and
survives until $t=1+\Delta$, the spike must have occurred in the time range
$[0,1]$.  Hence $Q_1=r e^{-rt}$.  For $k\geq 2$, the first spike must occur
within $[0,1]$, while the remaining (time-ordered) $k-1$ spikes can occur
anywhere within $[t_1,t]$.  To evaluate the integral in Eq.~(\ref{Qk}), we
may first integrate over $t_2,\ldots,t_k$.  This integral is
\begin{equation}
\label{simplex}
\int_{t_1}^t dt_2\, \int_{t_2}^t dt_3\, \cdots \int_{t_{k-1}}^t dt_k.
\end{equation}
The domain of the integral is a simplex of size $t-t_1$ and the integral is
just $(t-t_1)^{k-1}/(k-1)!$.  Finally, we integrate over the region $0<t_1<1$
to obtain
\begin{equation}
\label{Q1}
Q_k(t)={(rt)^k-(rt-r)^k\over k!}\, e^{-rt}.
\end{equation}
This expression actually holds for all $k\geq 0$.  {}From Eq.~(\ref{Q1}), we
then find that $S_1=1-e^{-r}$ in the time range $1<t<1+\Delta$.

Generally in the time range $1+(m-1)\Delta<t<1+m\Delta$, a tagged neuron
which has not fired must have experienced at least $m$ spikes; therefore
$Q_k=0$ for $k<m$.  To determine the non-zero $Q_k$'s -- those with $k\geq m$
-- note that the first $m$ spikes must obey the constraint of
Eq.~(\ref{range}), while each of the remaining $k-m$ spikes may lie anywhere
within the time interval $t_{m}$ and $t$.  The latter condition again defines
a simplex of size $t-t_m$.  This gives the contribution
$(t-t_m)^{k-m}/(k-m)!$ for the integral over these variables.  By this
reduction, the $k$-fold integral in Eq.~(\ref{Qk}) collapses to the $m$-fold
integral
\begin{equation}
\label{Ql}
Q_k(t)=r^k\,e^{-rt}\int 
{(t-t_m)^{k-m}\over (k-m)!}\prod_{j=1}^m dt_j. 
\end{equation}
In particular, for $k=m$, we may write $Q_m(t)$ as
$r^m\,e^{-rt}\,T_m(\Delta)$, where
\begin{equation}
\label{Tm}
T_m(\Delta)=\int_0^1 dt_1\int_{t_1}^{1+\Delta} dt_2\ldots 
\int_{t_{m-1}}^{1+(m-1)\Delta} dt_m.
\end{equation}
Remarkably, this expression has the simple closed-form representation (see
Appendix)
\begin{equation}
\label{T}
T_m(\Delta)={(1+m\Delta)^{m-1}\over m!}
\end{equation}
so that
\begin{equation}
\label{Qk=m}
Q_m(t)={r^m\,(1+m\Delta)^{m-1}\over m!}\,e^{-rt}.
\end{equation}
For large $m$ and also for$k>m$, the explicit expressions for $Q_k$ become
quite cumbersome; however, they are not needed to determine the asymptotics
of the survival probability.

We now use our result for $Q_k$ to determine the survival probability.  For
the time range $[1+(m-1)\Delta,1+m\Delta]$, we substitute Eq.~(\ref{Ql}) in
$S(t)=\sum_{k\geq m}Q_k(t)$ and perform the sum over $k$ to give
\begin{equation}
\label{S}
S_m=r^m\int e^{-rt_m}\prod_{j=1}^m dt_j. 
\end{equation}
This neat expression formally shows that the survival probability is constant
but $m$ dependent inside the time interval $[1+(m-1)\Delta,1+m\Delta]$.
These properties justify the notation in Eq.~(\ref{S}).

The integral on the right-hand side of Eq.~(\ref{S}) can be simplified by
integrating over $t_m$ to give
\begin{equation}
\label{rec}
S_m=S_{m-1}-r^{m-1}T_{m-1}\,e^{-r[1+(m-1)\Delta]}.
\end{equation}
This recursion relation allows us to express the survival probability in
terms of the $T_n$'s with $n<m$:
\begin{eqnarray*}
S_m=1-\sum_{n=0}^{m-1}r^n\, T_n\,e^{-r(1+n\Delta)}.
\end{eqnarray*}
Since $S_\infty=0$, we rewrite this as
\begin{equation}
\label{Sm}
S_m=\sum_{n=m}^\infty r^n\, T_n\,e^{-r(1+n\Delta)}
\end{equation}
which is more convenient for determining asymptotic behavior.

Let us first use this survival probability to compute the average time
interval $\langle t\rangle$ between consecutive firings of the same neuron.
This is
\begin{eqnarray}
\label{tav} 
\langle t\rangle &=&\int_0^\infty t\left(-{dS\over dt}\right) dt\nonumber\\
                 &=&\int_0^\infty S(t)\,dt\nonumber \\
                 &=&1+\Delta\sum_{m\geq 1} S_m\nonumber\\
                 &=&1+\Delta\sum_{n\geq 1}
                 n\,{r^n\, T_n\,e^{-r(1+n\Delta)}}.
\end{eqnarray}
In the first line, we use the fact that $-\dot S$ is just the probability
that the neuron fires at a time $t$ after its previous firing, and the last
line was derived by employing Eq.~(\ref{Sm}).  As discussed previously, the
average time between firings of the same neuron is $\langle
t\rangle=1+K\Delta$.  Equation (\ref{tav}) agrees with this iff the identity
\begin{equation}
\label{id2}
\sum_{n\geq 0} nT_n\,z^n={r\,e^r\over 1-r\Delta} 
\quad {\rm with}\quad z\equiv r\,e^{-r\Delta}
\end{equation}
holds.  This is also verified in Appendix.

Let us now interpret our results in the context of biological applications.
Typically, the number of neighboring neurons $K$ is large while the
spike-induced voltage decrement $\Delta$ of a neuron is small, so that the
total voltage decrease $K\Delta$ is of order one.  In other words, the limit
\begin{eqnarray*}
K\to\infty, \quad \Delta\to 0, \quad K\Delta={\cal O}(1)
\end{eqnarray*}
appears biologically relevant.  In this limit and when the time $t=1+m\Delta$
is large, the series for $S_m$ in Eq.~(\ref{Sm}) is geometric.  Hence, apart
from a prefactor, the survival probability is given by the first term in this
series:
\begin{equation}
\label{S3}
S_m\propto r^m\, T_m\,e^{-r(1+m\Delta)}.
\end{equation}
Using Eqs.~(\ref{S3}), (\ref{T}), and Stirling's formula, we deduce that
$S(t)$ decays exponentially with time, with the relaxation time in
Eq.~(\ref{exp}) given by
\begin{equation}
\label{tau}
\tau=-{\Delta\over P_1+\ln(1-P_1)}. 
\end{equation}

The different behavior of the two basic time scales, $\tau$ and $\langle
t\rangle=1/P_1$, is characteristic of biased diffusion near an absorbing
boundary in one dimension\cite{redner}.  Here, the mean survival time is
simply the distance from the particle to the absorbing boundary divided by
the mean velocity $v$.  In contrast, the survival probability asymptotically
decays as $\exp(-v^2t/D)$, so that $\tau=D/v^2$, independent of initial
distance.

It is instructive to interpret these results for our neural network, where a
single neuron can be viewed as undergoing a random walk in voltage, with a
step to smaller voltage of magnitude $\Delta$ occurring with probability
$r\,dt$ in a time interval $dt$ and a step to larger voltage of magnitude
$dt$ occurring with probability $1-r\,dt$.  For this random walk, the bias
velocity is $v=1-r\Delta=(1+K\Delta)^{-1}=P_1$.  This then reproduces
$\langle t\rangle=1/v=1+K\Delta=1/P_1$.  Moreover, the diffusion coefficient of
this random walk is simply $D\propto r\Delta^2$.  This random walk
description should be valid when $K\Delta\approx 1/P_1\gg 1$ or $r\Delta\to
1$, so that a tagged neuron experiences many spikes between firing events.
This then leads to $\tau=D/v^2\propto K^2\Delta^3$.  In this diffusive limit
of $P_1\to 0$, the limiting behavior of Eq.~(\ref{tau}) agrees with this
expression for $\tau$.

\section{Summary} 
 
We have determined the dynamical behavior of an integrate-and-fire neural
network in which there is purely inhibitory annealed coupling between
neighboring neurons.  The same behavior is also exhibited by a model with
quenched coupling.  Our model should be regarding as a ``toy'', since so many
realistic physiological features have been neglected.  However, this toy
model has the advantage of being analytically tractable.  We have determined
both the steady state properties of the network, as well as the complete
time-dependent behavior of a single neuron.  The latter gives rise to an
appealing first-passage problem for the probability for a neuron to survive a
time $t$ after its last firing.  This survival probability is piecewise
constant but with an overall exponential decay in time.

Given the simplicity of the model, it should be possible to incorporate some
of the more important features of real inhibitory neural networks, such as
neurons with leaky voltages and finite propagation velocity for inhibitory
signals, into the rate equation description.  These generalizations may
provide a tractable starting point to investigate more complex dynamical
behavior which are often the focus of neural network studies, such as
large-scale oscillations and macroscopic synchronization.

\medskip We thank C. Borgers, S. Epstein, N. Kopell, and S.~Yeung
for stimulating discussions.  We are also grateful to NSF grant
No.~DMR9978902 for partial financial support.

\appendix

\section{Basic Identities}

We use Eq.~(\ref{Sm}) to derive the identity (\ref{T}).  First, we note that
$S_0=1$.  By substituting this into Eq.~(\ref{Sm}) we obtain
\begin{equation}
\label{id1}
\sum_{n\geq 0} T_n\,z^n=e^r \quad {\rm with}\quad z\equiv r\,e^{-r\Delta}.
\end{equation}
The requirement that Eq.~(\ref{id1}) holds for arbitrary $r$ and $\Delta$
leads to unique set of $T_n$'s. To determine these $T_n$'s, let us treat $z$
as a complex variable. Then employing the Cauchy formula yields
\begin{eqnarray*}
T_n&=&{1\over 2\pi i}\oint{e^r\over z^{n+1}}\,dz\\
   &=&{1\over 2\pi i}\oint{e^r z'(r)\over [z(r)]^{n+1}}\,dr\\
   &=&{1\over 2\pi i}\oint{(1-r\Delta)\, e^{r(1+n\Delta)}
                            \over r^{n+1}}\,dr\\
   &=&{(1+n\Delta)^n\over n!}-\Delta\,{(1+n\Delta)^{n-1}\over (n-1)!}\\
   &=&{(1+n\Delta)^{n-1}\over n!}.
\end{eqnarray*}
 
Next we verify Eq.~(\ref{id2}) by repeating the procedure which has just been
used to check Eq.~(\ref{id1}).  As above, the quantity $nT_n$ may be written
in the integral representation
\begin{eqnarray*}
nT_n&=&{1\over 2\pi i}\oint{r\,e^r\over 1-r\Delta}\,{dz\over z^{n+1}}\\
&=&{1\over 2\pi i}\oint{r\,e^r z'(r)\over (1-r\Delta)[z(r)]^{n+1}}\,dr\\
   &=&{1\over 2\pi i}\oint{e^{r(1+n\Delta)}
                            \over r^n}\,dr\\
   &=&{(1+n\Delta)^{n-1}\over (n-1)!}.
\end{eqnarray*}

\end{multicols}
\end{document}